\documentclass[reprint, amsmath,amssymb,showpacs,floatfix, aps, prb, twocolumn, superscriptaddress]{revtex4-1}
\usepackage{graphicx}
\usepackage{dcolumn}
\usepackage{bm}
\usepackage{color}
\usepackage{hyperref}
\hypersetup{colorlinks=true,citecolor=blue,linkcolor=blue, urlcolor=blue}
\newcolumntype{M}[1]{>{\centering\arraybackslash}m{#1}}
\newcolumntype{N}{@{}m{0pt}@{}}

\bibpunct{[}{]}{,}{n}{}{}

\usepackage{amsfonts,amssymb,amsmath}
\usepackage[T1]{fontenc}
\usepackage{subcaption}
\usepackage{cancel}
\usepackage{xcolor}
\usepackage{graphicx}
\usepackage{tikz}
\usetikzlibrary{arrows,snakes,backgrounds}

\newcommand{\ZZ}{{\mathbb Z}}

\newcommand{\eps}{\epsilon}
\newcommand{\g}{\gamma}

\def\~#1{\widetilde{#1}}
\def\_#1{_{\mathrm{#1}}}

\begin{document}

\title{Bosonization in three spatial dimensions and a 2-form gauge theory}

\author{Yu-An Chen}
\email[E-mail: ]{ychen2@caltech.edu}
\affiliation {California Institute of Technology, Pasadena, CA 91125, USA}

\author{Anton Kapustin}
\affiliation {California Institute of Technology, Pasadena, CA 91125, USA}

\date{\today}

\begin{abstract}
\noindent  

We describe a 3d analog of the Jordan-Wigner transformation which maps an arbitrary fermionic system on a 3d spatial lattice to a 2-form $\ZZ_2$ gauge theory with an unusual Gauss law. An important property of this map is that it preserves the locality of the Hamiltonian. The map depends explicitly on the choice of a spin structure of the spatial manifold. We give examples of 3d bosonic systems dual to free fermions. We also describe the corresponding Euclidean lattice models, which is analogous to the generalized Steenrod square term in (3+1)D (compared to the Chern-Simon term in (2+1)D).
\end{abstract}
\maketitle

\section{Introduction and summary}

It is well known that every lattice fermionic system in 1d is dual to a lattice system of spins with a $\ZZ_2$ global symmetry (and vice versa). The duality is kinematic (independent of a particular Hamiltonian) and arises from the Jordan-Wigner transformation. Recently it has been shown that any lattice fermionic system in 2d is dual to a $\ZZ_2$ gauge theory with an unusual Gauss law \cite{CKR2017}. The fermion can be identified with the flux excitation of the gauge theory. The 2d duality is also kinematic. In this paper we extend these results to 3d systems. We show that every lattice fermionic system in 3d is dual to a $\ZZ_2$ 2-form gauge theory with an unusual Gauss law. Here ``2-form gauge theory'' means that the $\ZZ_2$ variables live on plaquettes, while the parameters of the gauge symmetry live on links. 2-form gauge theories in 3+1D have local flux excitations, and the unusual Gauss law ensures that these excitations are fermions.\footnote{In contrast, a 2-form $\ZZ_2$ gauge theory with the standard Gauss law is mapped, by a 3d version of the Kramers-Wannier duality, to a theory of bosonic spins.}

The form of the modified Gauss law is largely dictated by the observation first made in \cite{GK2016} that a bosonization of fermionic systems in $d$ dimensions must have a global $(d-1)$-form $\ZZ_2$ symmetry with a particular 't Hooft anomaly. The standard Gauss law leads to a trivial 't Hooft anomaly, so bosonization requires us to modify it in a particular way. The precise form of the modified Gauss law and the bosonization map depends on the choice of the lattice. We describe them in two cases: the cubic lattice and a 3d triangulation.

Our 3d bosonization map is kinematic and local in the same sense as the Jordan-Wigner map: every local bosonic observable on the fermionic side, including the Hamiltonian density, is mapped to a local gauge-invariant observable on the gauge theory side.

In the literature, there are examples of specific bosonic models in 3d with emergent fermions. Our general construction is reminiscent of the work by Levin and Wen \cite{LW2006}. These authors constructed systems of rotors which have emergent fermions. In our approach rotors are replaced with $\ZZ_2$ spins. There are also several proposals for an analog of the Jordan-Wigner map in arbitrary  dimensions \cite{BravyiKitaev,Ball,VersCirac}. Our construction is most similar to that of Bravyi and Kitaev \cite{BravyiKitaev}. One advantage of our construction is that we can clearly identify the kind  of 3d bosonic systems that are dual to fermionic systems: they possess global 2-form $\ZZ_2$ symmetry with a specific 't Hooft anomaly, as proposed in \cite{GK2016}. It is also manifest in our approach that the bosonization map depends on a choice of spin structure.

Our bosonization method allows for an easy construction of bosonic systems dual to free fermions with an arbitrary dispersion law. As an illustration, we describe a bosonic model on a cubic lattice whose dual fermionic description involves Dirac cones. It can be regarded as a 3d analog of the Kitaev honeycomb model. Other 3d analogs of the honeycomb model have been proposed in \cite{honeycomb3d1,honeycomb3d2}. We also identify some Euclidean bosonic 4D models which are dual to free fermions. These models can be understood as 2-form $\ZZ_2$ gauge theories whose action involves a topological term.

\section{Bosonization on a three-dimensional lattice}

\subsection{Cubic lattice}\label{sec:square}

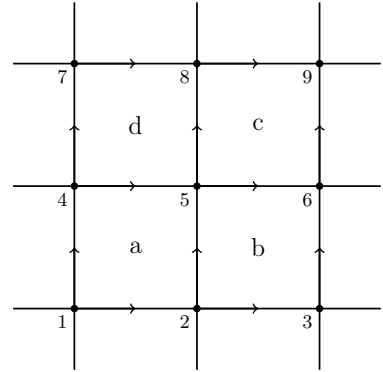
\begin{figure}
\centering
\resizebox{5cm}{!}{%
\begin{tikzpicture}
\draw[thick] (-3,0) -- (3,0);\draw[thick] (-3,-2) -- (3,-2);\draw[thick] (-3,2) -- (3,2);
\draw[thick] (0,-3) -- (0,3);\draw[thick] (-2,-3) -- (-2,3);\draw[thick] (2,-3) -- (2,3);
\draw[->] [thick](0,0) -- (1,0);\draw[->][thick] (0,2) -- (1,2);\draw[->][thick] (0,-2) -- (1,-2);
\draw[->][thick] (0,0) -- (0,1);\draw[->][thick] (2,0) -- (2,1);\draw[->][thick](-2,0) -- (-2,1);
\draw[->][thick] (-2,0) -- (-1,0);\draw[->][thick] (-2,2) -- (-1,2);\draw[->][thick](-2,-2) -- (-1,-2);
\draw[->][thick] (-2,-2) -- (-2,-1);\draw[->][thick] (0,-2) -- (0,-1);\draw[->] [thick](2,-2) -- (2,-1);
\filldraw [black] (-2,-2) circle (1.5pt) node[anchor=north east] {1};
\filldraw [black] (0,-2) circle (1.5pt) node[anchor=north east] {2};
\filldraw [black] (2,-2) circle (1.5pt) node[anchor=north east] {3};
\filldraw [black] (-2,-0) circle (1.5pt) node[anchor=north east] {4};
\filldraw [black] (0,0) circle (1.5pt) node[anchor=north east] {5};
\filldraw [black] (2,0) circle (1.5pt) node[anchor=north east] {6};
\filldraw [black] (-2,2) circle (1.5pt) node[anchor=north east] {7};
\filldraw [black] (0,2) circle (1.5pt) node[anchor=north east] {8};
\filldraw [black] (2,2) circle (1.5pt) node[anchor=north east] {9};
\draw (-1,-1) node{\large a};
\draw (1,-1) node{\large b};
\draw (1,1) node{\large c};
\draw (-1,1) node{\large d};
\end{tikzpicture}
}
\caption{Bosonization on a square lattice requires constraints on vertices.}
\label{fig:square}
\end{figure}

We begin by reviewing the 2d bosonization on a square lattice following \cite{CKR2017}. The set of vertices, edges, and faces are denoted $V,E,F$, and their elements $v,e,f$. On each face $f$ of the lattice we place a single pair of fermionic creation-annihilation operators $c_f,c_f^\dagger$, or equivalently a pair of Majorana fermions $\gamma_f,\gamma'_f$. The even fermionic algebra consists of local observables with a trivial fermionic parity (i.e. $(-1)^F=1$). It is generated by $(-1)^{F_f}=-i\gamma_f\gamma'_f$ and $S_e=i\gamma_{L(e)}\gamma'_{R(e)}$, where $L(e)$ and $R(e)$ are faces to the left and right of $e$, with respect to some orientation of $e$. 

The bosonic dual of this system involves $\ZZ_2$-valued spins on the edges of the square lattice. For every edge $e$ we define a unitary operator $U_e$ which squares to $1$. Labeling the faces and vertices as in Fig. \ref{fig:square}, we  define:
\begin{equation}
\begin{split}
U_{56} &= X_{56} Z_{25} \\
U_{58} &= X_{58} Z_{45} 
\end{split}
\end{equation}
where $X$, $Z$ are Pauli matrices acting on a spin at each edge. Operators $U_e$ for other edges are defined by using translation symmetry. 

It has been shown in \cite{CKR2017}  that $U_e$ and $S_e$ satisfy the same commutation relations. We also identify fermionic parity $(-1)^{F_f}$ at each face with the ``flux operator'' $W_f \equiv \prod_{e \subset f} Z_e$. The bosonization map is
\begin{equation}
\begin{split}
(-1)^{F_f} = -i \g_f \g^\prime_f &\longleftrightarrow W_f \\
 S_e =  i \g_{L(e)} \g^\prime_{R(e)} &\longleftrightarrow U_e.
\end{split}
\end{equation}
The condition $ (-1)^{F_a} (-1)^{F_c} S_{{58}} S_{{56}} S_{{25}} S_{{45}} = 1$ on fermionic operators gives gauge constraints $W_{f_c} \prod_{e \supset v_5} X_e  =1$ for bosonic operators, or generally
\begin{equation}
W_{\text{NE}(v)} \prod_{e \supset v} X_e = 1
\label{eq:gauge constraint at vertex}
\end{equation}
where $\text{NE}(v)$ is the face northeast of $v$. Eq.~(\ref{eq:gauge constraint at vertex}) is the modified Gauss law for a 2d gauge theory.

Next, we introduce our bosonization method on an infinite 3d cubic lattice. Suppose that we have a model with fermions living at the centers of cubes. Let us describe the generators and relations in the algebra of local observables with trivial fermion parity (the even fermionic algebra for short).

On each cube $t$ we have a single fermionic creation operator $c_t$ and a single fermionic annihilation operator $c^\dagger_t$ with the usual anticommutation relations. The fermionic parity operator on cube $t$ is $(-1)^{F_t}=(-1)^{c_t^\dagger c_t}$. It is a ``$\ZZ_2$ operator'' (i.e.\ it squares to $1$). All operators $(-1)^{F_t}$ commute with each other. The even fermionic algebra is generated by these operators and the operators $c_t^\dagger c_{t'}$, $c_t c_{t'}$, and their Hermitean conjugates, where $t$ and $t'$ are two cubes which share a face. Overall, we get four generators for each face and one generator for each cube. 
It is easier to work in the Majorana basis
\begin{equation}
\gamma_t = c_t+c_t^\dagger,\quad \gamma'_t=(c_t-c_t^\dagger)/i.
\end{equation}
The even fermionic algebra is generated by $i \g_{L(f)} \g^\prime_{R(f)}$ and $ -i \g_t \g^\prime_t$ where each face is assigned an orientation from cube $L(f)$ to cube $R(f)$.

\begin{figure}[htb]
\centering
\resizebox{8 cm}{!}{%
\begin{tikzpicture}
\draw[thick] (-0.5,0) -- (2.5,0);\draw[thick] (0.5,1) -- (3.5,1);\draw[thick] (0.5,2.5) -- (3.5,2.5);

\draw[thick] (-0.3,-0.3) -- (1.6,1.6);\draw[thick] (1.7,-0.3) -- (3.3,1.3);\draw[thick] (-0.3,1.2) -- (1.3,2.8);

\draw[thick] (0,-0.5) -- (0,2);\draw[thick] (1,0.15) -- (1,3);\draw[thick] (3,0.15) -- (3,3);

\draw[thick] (-0.5,1.5) -- (0.5,1.5); \draw[thick] (2,-0.5) -- (2,0.5);\draw[thick] (2.3,1.8) -- (3.3,2.8);

\draw[green] (0.15,-0.15) -- (1.15,0.85);
\draw[red] (1.15,0.85) -- (3.15,0.85);
\draw[blue] (1.15,0.85) -- (1.15,2.35);

\draw[->][thick] (-2,1) -- (-1,1);\draw[->][thick] (-2,1) -- (-2,2);\draw[->][thick] (-2,1) -- (-1.4,1.6);
\draw (-0.8,1) node{x};\draw (-1.2,1.8) node{y};\draw (-2,2.2) node{z};

\filldraw[white] (2.5,2) circle (2pt); \draw (2.5,2) circle (2pt);
\filldraw[white] (0.5,0.5) circle (2pt); \draw (0.5,0.5) circle (2pt) node[anchor=south] {2};
\filldraw[white] (0,0.75) circle (2pt); \draw (0,0.75) circle (2pt);
\filldraw[white] (1,1.75) circle (2pt); \draw (1,1.75) circle (2pt) node[anchor=north east] {1};
\filldraw[white] (3,1.75) circle (2pt); \draw (3,1.75) circle (2pt);
\filldraw[white] (0.5,2) circle (2pt); \draw (0.5,2) circle (2pt);
\filldraw[white] (2,2.5) circle (2pt); \draw (2,2.5) circle (2pt);

\filldraw[green] (1,0.25) circle (2pt);\draw (1,0.25) circle (2pt) node[anchor=south west] {8};
\filldraw[green] (1,0) circle (2pt); \draw (1,0) circle (2pt) node[anchor=north] {7};
\filldraw[red] (3,0.25) circle (2pt); \draw (3,0.25) circle (2pt) node[anchor=west] {6};
\filldraw[red] (2.5,0.5) circle (2pt); \draw (2.5,0.5) circle (2pt) node[anchor=north] {5};
\filldraw[blue] (1.5,1.5) circle (2pt); \draw (1.5,1.5) circle (2pt) node[anchor=south west] {4};
\filldraw[blue] (2,1) circle (1.5pt); \draw (2,1) circle (2pt) node[anchor=south] {3};
\end{tikzpicture}
}
\caption{(Color online) For edges in the dual lattice, the "framing" is defined by green, red, and blue edges, which is a small shift of duel edges \cite{LW2006}. Given a dual edge $f$, the operator $U_f$ is defined as $X_f$ times $Z_{f'}$ for those $f'$ which intersect the framing of $f$ when projected to the plane (i.e. $U_{f_1}=X_1 Z_3 Z_4,$ $U_{f_2}=X_2 Z_7 Z_8,$ and $U_{f_3}=X_3 Z_5 Z_6$  ). }
\label{fig:cube1}
\end{figure}
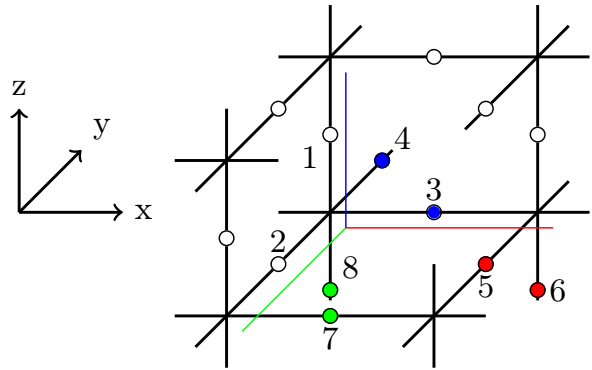

To illustrate the definition of these operators, we draw the dual lattice of the original lattice. In Fig. \ref{fig:cube1}, fermions live on vertices and the orientations of each dual edge (face of the original lattice) are taken to be along $+x$, $+y$, and $+z$ directions.
The Majorana hopping operator is defined by $S_f= i \g_{L(f)} \g^\prime_{R(f)}$ where $L(f)$ and $R(f)$ are source and sink (starting and ending points) of dual edge $f$ in the dual lattice. $S_{f_i}$ and $S_{f_j}$ anti-commute only when both dual edges $f_i$ and $f_j$ start from the same point or both end at the same point.

The dual bosonic system has $\ZZ_2$ spins living on faces of the original lattice (or equivalently, on edges of the dual lattice). To define bosonic hopping operators $U_f$, we need to choose a framing for each edge of the dual lattice, i.e. a small shift of each  dual edge along some orthogonal direction. We also assume that when projected on some generic plane (such as the plane of the page) a shifted dual edge intersects all dual edges transversally. For example, in Fig. \ref{fig:cube1} such a framing is indicated by red, green and blue lines (for dual edges along $x$, $y$ and $z$ directions, respectively), and the shift of the dual edge $1$ intersects dual edges $3$ and $4$.\footnote{There are many choices of framing, and accordingly many versions of the bosonization map. By construction, they are related by   automorphisms of the algebra of observables.} Now we define $U_f$ as a product of $X_f$ with all $Z_{f^\prime}$ such that $f^\prime$ intersects the framing of $f$ when projected to the plane of the page. For example, the hopping operator for the dual edge $1$ is  $U_1=X_1 Z_3 Z_4$. Notice that $U_1$, $U_3$, and $U_4$ anti-commute with each other and $U_3$, $U_5$, and $U_6$ anti-commute with each other, while $U_2$ and $U_3$ commute, and $U_1$ and $U_8$ commute. 

One can check that $S_f$ and $U_f$ have the same commutation relations. Therefore, the bosonization map in 3D can be defined as follows:

\begin{enumerate}

\item
For any cube $t$ let $W_t \equiv \prod_{f \subset t} Z_f$. We identify the fermionic states $ |F_t=0 \rangle$ and $ |F_t=1 \rangle$  with bosonic states for which $W_t=1$ and $W_t=-1$, respectively. Thus
\begin{equation}
(-1)^{F_t} = -i \g_t \g^\prime_t \longleftrightarrow W_t.
 \label{eq:bosonization assumption 1}
\end{equation}

\item
The fermionic hopping operator $S_f$ is identified with $U_f$ defined above:
\begin{equation}
 S_f =  i \g_{L(f)} \g^\prime_{R(f)} \longleftrightarrow U_f.
 \label{eq:bosonization assumption 2}
\end{equation}
\end{enumerate}

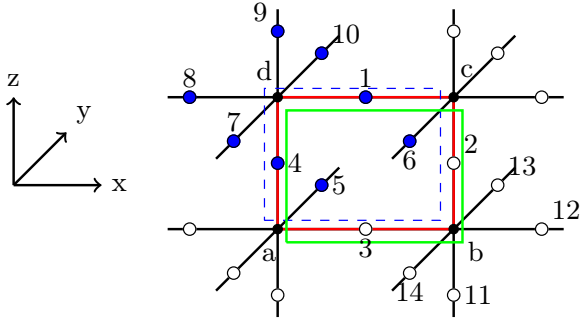
\begin{figure}[htb]
\centering
\resizebox{8 cm}{!}{%
\begin{tikzpicture}
\draw[thick] (-1.25,0) -- (3.25,0);\draw[thick] (-1.25,1.5) -- (3.25,1.5);

\draw[thick] (0,-1) -- (0,2.5);\draw[thick] (2,-1) -- (2,2.5);

\draw[thick] (-0.7,-0.7) -- (0.7,0.7);\draw[thick] (1.3,-0.7) -- (2.7,0.7);\draw[thick] (-0.7,0.8) -- (0.7,2.2);\draw[thick] (1.3,0.8) -- (2.7,2.2);

\draw[red,thick] (0,0) -- (2,0) -- (2,1.5) -- (0,1.5) -- (0,0);
\draw[blue,dashed] (-0.15,0.1) -- (1.85,0.1) -- (1.85,1.6) -- (-0.15,1.6) -- (-0.15,0.1);
\draw[green,thick] (0.1,-0.15) -- (2.1,-0.15) -- (2.1,1.35) -- (0.1,1.35) -- (0.1,-0.15);

\filldraw[blue] (1,1.5) circle (2pt);\draw (1,1.5) circle (2pt) node[anchor=south] {1};
\filldraw[blue] (0,0.75) circle (2pt);\draw (0,0.75) circle (2pt) node[anchor=west] {4};
\filldraw[blue] (0.5,0.5) circle (2pt);\draw (0.5,0.5) circle (2pt) node[anchor=west] {5};
\filldraw[blue] (1.5,1) circle (2pt);\draw (1.5,1) circle (2pt) node[anchor=north] {6};
\filldraw[blue] (-0.5,1) circle (2pt);\draw (-0.5,1) circle (2pt) node[anchor=south] {7};
\filldraw[blue] (-1,1.5) circle (2pt);\draw (-1,1.5) circle (2pt) node[anchor=south] {8};
\filldraw[blue] (0,2.25) circle (2pt);\draw (0,2.25) circle (2pt) node[anchor=south east] {9};
\filldraw[blue] (0.5,2) circle (2pt);\draw (0.5,2) circle (2pt) node[anchor=south west] {10};

\filldraw[white] (2.5,2) circle (2pt); \draw (2.5,2) circle (2pt);
\filldraw[white] (2,2.25) circle (2pt); \draw (2,2.25) circle (2pt);
\filldraw[white] (3,1.5) circle (2pt); \draw (3,1.5) circle (2pt);
\filldraw[white] (-1,0) circle (2pt); \draw (-1,0) circle (2pt);
\filldraw[white] (-0.5,-0.5) circle (2pt); \draw (-0.5,-0.5) circle (2pt);
\filldraw[white] (0,-0.75) circle (2pt); \draw (0,-0.75) circle (2pt);
\filldraw[white] (2,0.75) circle (2pt); \draw (2,0.75) circle (2pt) node[anchor=south west] {2};
\filldraw[white] (1,0) circle (2pt); \draw (1,0) circle (2pt) node[anchor=north] {3};
\filldraw[white] (2,-0.75) circle (2pt); \draw (2,-0.75) circle (2pt) node[anchor=west] {11};
\filldraw[white] (3,0) circle (2pt); \draw (3,0) circle (2pt) node[anchor=south west] {12};
\filldraw[white] (2.5,0.5) circle (2pt); \draw (2.5,0.5) circle (2pt) node[anchor=south west] {13};
\filldraw[white] (1.5,-0.5) circle (2pt); \draw (1.5,-0.5) circle (2pt) node[anchor=north] {14};

\draw[->][thick] (-3,0.5) -- (-2,0.5);\draw[->][thick] (-3,0.5) -- (-3,1.5);\draw[->][thick] (-3,0.5) -- (-2.4,1.1);
\draw (-1.8,0.5) node{x};\draw (-2.2,1.3) node{y};\draw (-3,1.7) node{z};
\filldraw[black] (0,0) circle (1.5pt);\draw (-0.1,-0.3) node{a};
\filldraw[black] (2,0) circle (1.5pt);\draw (2.25,-0.25) node{b};
\filldraw[black] (2,1.5) circle (1.5pt);\draw (2.15,1.8) node{c};
\filldraw[black] (0,1.5) circle (1.5pt);\draw (-0.15,1.8) node{d};

\end{tikzpicture}
}
\caption{(Color online) The framing of the hopping term defined previously is indicated by the green square, while the gauge constraint involves the $Z$ operators in the opposite framing (blue dashed square).}
\label{fig:cube2}
\end{figure}

As in 2d, the bosonic operators satisfy some constraints. In Fig. \ref{fig:cube2}, we calculate the product of $S_f$ around the red square on the dual lattice:

\begin{equation}
\begin{split}
 &S_{f_1} S_{f_2} S_{f_3} S_{f_{4}} \\
 =& (i \g_d \g^\prime_c)  (i \g_b \g^\prime_c)  (i \g_a \g^\prime_b)  (i \g_a \g^\prime_d) \\
=& - (-i \g_b \g^\prime_b ) (-i \g_d \g^\prime_d ) \\
=& - (-1)^{F_b} (-1)^{F_d} \longleftrightarrow - W_b W_d.
\end{split}
\label{eq: S delta cube}
\end{equation}
Its bosonic dual defined by \eqref{eq:bosonization assumption 2} is the product of the corresponding operators $U_f$. Their definition involves a framing of the red square given by the green square:
\begin{equation}
\begin{split}
 &U_{f_1} U_{f_2} U_{f_3} U_{f_4} \\
 =& (X_1 Z_2 Z_6) (X_2 Z_{12} Z_{13}) (X_3 Z_{11} Z_{14}) (X_4 Z_3 Z_5) \\
=& - X_1 X_2 X_3 X_4 Z_5 Z_6 (Z_2 Z_3 Z_{11} Z_{12} Z_{13} Z_{14}) \\
=& - X_1 X_2 X_3 X_4 Z_5 Z_6 W_b.
\end{split}
\label{eq: U delta cube}
\end{equation}
Comparing \eqref{eq: S delta cube} and \eqref{eq: U delta cube}, we get the constraint
\begin{equation}
\begin{split}
1 = &X_1 X_2 X_3 X_4 Z_5 Z_6 W_d \\
= &X_1 X_2 X_3 X_4 Z_1 Z_4 Z_5 Z_6 Z_7 Z_8 Z_9 Z_{10}
\end{split}
\label{eq:cubic gauge constraint}
\end{equation}
The operators $Z$'s are the edges crossed by dashed square in Fig. \ref{fig:cube2}. The framing for gauge constraints is opposite to the framing used to define hopping operators. We have a gauge constraint for each face of dual lattice. In terms of the original lattice, there is one gauge constraint for each edge. All these constraints commute and thus define a $\ZZ_2$ 2-form gauge theory with an unusual Gauss law.

\subsection{Examples}
\subsubsection{Soluble 3+1D lattice gauge theories}

The standard Gauss law for a 2-form $\ZZ_2$ gauge theory is $\prod_{f\supset e} X_f=1$. Such a bosonic gauge theory is dual to a theory of bosonic spins living on the vertices of the dual lattice. In particular, the quantum Ising model can be described by a $\ZZ_2$ 2-form gauge theory with the Hamiltonian
\begin{equation}\label{Ising}
H_{Ising}=g^2 \sum_f X_f + \frac{1}{g^2} \sum_t W_t.
\end{equation}
This model is not soluble. 

If we impose the modified Gauss law (\ref{eq:cubic gauge constraint}) instead, the simplest analogous  gauge-invariant Hamiltonian is
\begin{equation}
H_b= g^2 \sum_f U_f + \frac{1}{g^2} \sum_t W_t.
\end{equation}
The first and second term can be thought of as the kinetic and potential energies, respectively. This is dual to the fermionic Hamiltonian
\begin{equation}
    \begin{split}
        H_f =& t \sum_f \Big( c_{L(f)} c_{R(f)} - c^\dagger_{L(f)} c^\dagger_{R(f)} \\
        &+ c^\dagger_{L(f)} c_{R(f)} + c^\dagger_{R(f)} c_{L(f)} \Big) + \mu \sum_t c^\dagger_t c_t
    \end{split}
\end{equation}
where $t=g^2$ and $\mu=\frac{2}{g^2}$. The fermionic  Hamiltonian is free and thus soluble.
By Fourier transform $c_{\vec{x}} = \frac{1}{\sqrt{N}} \sum_{\vec{k}} e^{i \vec{k} \cdot \vec{x}} c_{\vec{k}}$, the fermionic Hamiltonian becomes
\begin{equation}
H_f = \sum_{\vec{k}} \eps_{\vec{k}} c^\dagger_{\vec{k}} c_{\vec{k}} + \sum_{\vec{k}} (\Delta_{\vec{k}} c_{\vec{k}} c_{-\vec{k}}+\mathrm{h.c.})
\label{eq:before BdG}
\end{equation}
with $\eps_{\vec{k}}= \mu + 2t (\cos k_x + \cos k_y + \cos k_z)$ and $\Delta_{\vec{k}} = t (e^{- i k_x} + e^{- i k_y} + e^{- i k_z})$. The Hamiltonian \eqref{eq:before BdG} can be written in the Bogoliubov-de-Gennes (BdG) formalism as \begin{equation}
H_f=\frac{1}{2} \sum_{\vec{k}} \Psi^\dagger_{\vec{k}} H_{\_{BDG}}(\vec{k}) \Psi_{\vec{k}}
\end{equation}
with
\begin{equation}
H_{\_{BDG}}(\vec{k})=
\begin{bmatrix}
\eps_{\vec{k}} & -\Delta^*_{\vec{k}}  \\
-\Delta_{\vec{k}} & -\eps_{\vec{k}}  \\

\end{bmatrix}
,\quad
\Psi_{\vec{k}}=
\begin{bmatrix}
c_{\vec{k}} \\ c^\dagger_{-\vec{k}} 
\end{bmatrix}.
\end{equation}
The spectrum is 
\begin{equation}
\begin{split}
&E^2 = \\
& \, t^2(3+ 2 \cos(k_x - k_y) + 2 \cos(k_x - k_z) + 2 \cos(k_y - k_z)) \\
&+ [\mu + 2t (\cos k_x + \cos k_y + \cos k_z)]^2.
\end{split}
\end{equation}
Notice that for $\mu=0$ the gap closes for $\vec{k}=(q,q+\frac{2\pi}{3},q+\frac{4\pi}{3})$ for arbitrary $q$. 

\subsubsection{Bosonic model with Dirac cones}

Using the bosonization map \eqref{eq:bosonization assumption 1} and \eqref{eq:bosonization assumption 2}, we can construct an equivalent bosonic model for any arbitrary fermionic model. For instance, Ref. ~\cite{Creutz} constructs a fermionic model on a cubic lattice with Dirac cones:
\begin{equation}
H = -t \sum_{\vec{r}} (s_x(\vec{r}) c^\dagger_{\vec{r}+\hat{x}} c_{\vec{r}} + s_y(\vec{r}) c^\dagger_{\vec{r}+\hat{y}} c_{\vec{r}} + s_z(\vec{r}) c^\dagger_{\vec{r}+\hat{z}} c_{\vec{r}} + \text{h.c.})
\label{eq: Creutz Hamiltonian}
\end{equation}
with $s_x(\vec{r})$, $s_y(\vec{r})$, and $s_z(\vec{r})$ defined as
\begin{equation}
\begin{split}
s_x(i,j,k) &= 1 \\
s_y(i,j,k) &= (-1)^i \\
s_z(i,j,k) &= (-1)^{i+j}.
\end{split}
\end{equation}
It is a model with nearest neighbor hopping. The  spectrum is
\begin{equation}
E = \pm 2t \sqrt{\cos^2 k_x +\cos^2 k_y +\cos^2 k_z}
\end{equation}
with two Dirac cones at $\vec{k}=(\pi/2,\pi/2,\pi/2)$ and $\vec{k}=(3\pi/2,\pi/2,\pi/2)$.
Applying the bosonization map, the corresponding bosonic Hamiltonian is
\begin{equation}
\begin{split}
H =& - \frac{t}{2} \sum_{f_x} s_x(L(f_x)) U_{f_x} (1- W_{L(f_x)} W_{R(f_x)}) \\
& - \frac{t}{2} \sum_{f_y} s_y(L(f_y)) U_{f_y} (1- W_{L(f_y)} W_{R(f_y)}) \\
& - \frac{t}{2} \sum_{f_z} s_z(L(f_z)) U_{f_z} (1- W_{L(f_z)} W_{R(f_z)}),
\end{split}
\label{eq:bosonic Creutz model}
\end{equation}
where $f_x$, $f_y$, $f_z$ refer to faces normal to $x$, $y$, $z$-directions, with gauge constraints \eqref{eq:cubic gauge constraint}. On the bosonic side, it is very nontrivial to see that the model describes  Dirac cones.

\subsection{Triangulation}

\begin{figure}[htb]
\centering
\resizebox{5cm}{!}{%
\begin{tikzpicture}[scale=0.6]
\draw[->] [cyan] (1.5,0.7) -- (4.5,0.7) -- (2.2,2.425) -- (1,0.625);
\draw[->] [cyan] (4,3) -- (2.75,2.6875) -- (4.8,1.15) -- (4.1, 2.375);
\draw[thick] (0,0) -- (6,0);\draw[thick] (0,0) -- (2,3);\draw[thick] (0,0) -- (4,3.5);
\draw[thick] (6,0) -- (4,3.5);\draw[thick] (2,3) -- (4,3.5);\draw[dashed,thick] (2,3) -- (6,0);
\draw[->] [thick] (0,0) -- (3,0);\draw[->] [thick](0,0) -- (2,1.75);\draw[->] [thick](2,3) -- (1,1.5);
\draw[->] [thick] (6,0) -- (5,1.75);\draw[->] [thick](2,3) -- (3,3.25);\draw[->] [dashed,thick](2,3) -- (4,1.5);
\draw[->] [red] (4,3) -- (1,0.375) -- (5.5,0.375) -- (4.5,2.125);
\draw[->] [red] (3,2.875) -- (2,2.625) -- (1,1.125) -- (2.5,2.4374);

\draw (2,3)  node[anchor=south east] {0};
\draw (0,0)  node[anchor=north east] {1};
\draw (6,0)  node[anchor=north west] {2};
\draw (4,3.5)  node[anchor=south west] {3};
\end{tikzpicture}
}
\caption{(Color online) A branching structure on a tetrahedron. The orientation of each face is determined by the right-hand rule. We defined this as the ``$+$'' tetrahedron, the directions of faces $012$ and $023$ are inward (blue) while the directions of faces $123$ and $013$ are outward (red). The directions of faces are reversed in the ``$-$'' tetrahedron (mirror image of this tetrahedron).}
\label{fig:tetrahedron}
\end{figure}
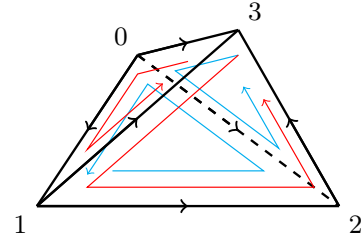

The bosonization method described above also works for any triangulation. For an arbitrary triangulation $T$ of a 3d manifold $M$, we choose a branching structure. A branching structure is a choice of an orientation on each edge such that there is no oriented loop on any triangle. One simple way is to label vertices by different numbers and assign the direction of an edge from the vertex with smaller number to the vertex with larger number (see Fig. \ref{fig:tetrahedron}). Each tetrahedron has two inward faces and two outward faces (by right-hand rule). We place fermions at the centers of tetrahedrons. Each tetrahedron $t$ contains Majorana operators $\g_t$ and $\g^\prime_t$. We define the fermionic hopping operator on each face $f$ as
\begin{equation}
S_f = i \g_{L(f)} \g^\prime_{R(f)}
\end{equation}
where $L(f)$/$R(f)$ is the tetrahedron with $f$ as a outward/inward face. Notice that $S_f$ and $S_{f^\prime}$ anti-commute only when $f$ and $f^\prime$ share a tetrahedron with both $f$ and $f^\prime$ inward or outward. To express this property mathematically, we introduce (higher) cup product used in algebraic topology. The definition and properties of the (higher) cup products are reviewed in Appendix \ref{sec:cup product}.
If $\beta_1$ and $\beta_2$ are 2-cochains, then
\begin{equation}
\beta_1 \cup_1 \beta_2 (0123) = \beta_1 (023) \beta_2 (012) + \beta_1(013) \beta_2(123).
\end{equation}
Therefore, the commutation relations can be expressed as
\begin{equation}
S_f S_{f^\prime} = (-1)^{\int f \cup_1 f^\prime + f^\prime \cup_1 f} S_{f^\prime} S_f
\label{eq:Sf commutation relation}
\end{equation}
where we abuse the notation $f \in C^2(T,\ZZ_2)$ as a 2-cochain with value 1 on face $f$ and 0 otherwise.
The even fermionic algebra is generated by the operators $S_f$ for all faces and the fermionic parity operators $(-1)^{F_t}$ for all tetrahedra.

The dual bosonic variables are $\ZZ_2$ spins which live on faces of the triangulation. As before, the flux operator
$$
W_t=\prod_{f\supset t} X_f
$$
corresponds to $(-1)^{F_t}$ under the bosonization map.

Next we need to find bosonic operators $U_f$ which have the same commutation relation as fermionic operators $S_f$. We should define $U_f$ as $X_f$ times $Z_{f^\prime}$ for some faces $f^\prime$ which share a tetrahedron with $f$ and have the same orientation with respect to the tetrahedron. One way to define $U_f$ is
\begin{equation}
U_f = X_f \prod_{t \in \{ L(f),R(f)\}} Z_{t_{023}}^{f(t_{012})} Z_{t_{013}}^{f(t_{123})} = X_f \prod_{f^\prime} Z_{f^\prime}^{\int f^\prime \cup_1 f}.
\label{eq:hopping defined by cup1}
\end{equation}
$U_f$ satisfy the commutation relation
\begin{equation}
U_f U_{f^\prime} = (-1)^{\int f \cup_1 f^\prime + f^\prime \cup_1 f} U_{f^\prime} U_f
\end{equation}
which is the same as \eqref{eq:Sf commutation relation}.

The final step is to determine the constraints on bosonic variables. There is one such constraint for  each edge $e$. In the product $ \prod_{f \supset e} S_f$, the only surviving terms are $-i \g_t \g^\prime_t$ with one face going inward and one face going outward of $t$. The term $-i \g_t \g^\prime_t$ is bosonized to $W_t \equiv \prod_{f \subset t} Z_f$. Therefore, the product can be written as
\begin{equation}
\prod_{f \supset e} S_f \sim \prod_{t | e=t_{01},t_{03},t_{12},t_{23}} W_t
\label{eq:prod Sf1}
\end{equation}
where $\sim$ means that it is equal up to a sign, which will be treated carefully in the next paragraph.
For a tetrahedron $t$ containing an edge $e$ with adjacent faces $f_1$ and $f_2$, consider the following product which gives $W_t$ for $e=t_{01},t_{03},t_{12},t_{23}$ and $1$ otherwise:

\begin{equation}
    \begin{split}
        &Z_{f_1} Z_{f_2} \prod_{f^\prime \subset t} Z_{f^\prime}^{(f_1+f_2) \cup_1 f^\prime + f^\prime \cup_1 (f_1+f_2)}\\
        &= 
        \begin{cases}
            W_t, & \text{if } e=t_{01},t_{03},t_{12},t_{23} \\
            1,   & \text{otherwise}
        \end{cases}
    \end{split}
\label{eq:math way to express prod Sf1}
\end{equation}
Substituting this into \eqref{eq:prod Sf1}, we have
\begin{equation}
\prod_{f \supset e} S_f \sim \prod_{f \supset e} \prod_{f^\prime} Z_{f^\prime}^{\int f^\prime \cup_1 f + f \cup_1 f^\prime} = \prod_{f^\prime} Z_{f^\prime}^{\int f^\prime \cup_1 \delta e + \delta e \cup_1 f^\prime}
\label{eq:prod Sf2}.
\end{equation}
On the other hand, the product $\prod_{f \supset e} U_f$ is
\begin{equation}
\prod_{f \supset e} U_f \sim \prod_{f \supset e} X_f \prod_{f^\prime} Z_{f^\prime}^{\int f^\prime \cup_1 f} = ( \prod_{f \supset e} X_f ) \prod_{f^\prime} Z_{f^\prime}^{\int f^\prime \cup_1 \delta e}
\label{eq:prod Uf}.
\end{equation}
Identifying \eqref{eq:prod Sf2} and \eqref{eq:prod Uf} gives
\begin{equation}
(\prod_{f \supset e} X_f) \prod_{f^\prime} Z_{f^\prime}^{\int  \delta e \cup_1 f^\prime } = 1
\label{eq:gauge constraints cup1}
\end{equation}
This is the modified Gauss law (gauge constraint) on each edge $e$. One can check that constraints for  different edges $e_1$ and $e_2$ commute since
\begin{multline}
\int \left(\delta e_1 \cup_1 \delta e_2 + \delta e_2 \cup_1 \delta e_1\right) =\\
=\int \left(e_1 \cup \delta e_2 + \delta e_2 \cup e_1 + e_2 \cup \delta e_1 + \delta e_1 \cup e_2\right) = 0
\label{eq:use cup1 1}
\end{multline}
where we have used the property $\int \delta e_1 \cup_1 \delta e_2 = \int \left(e_1 \cup \delta e_2 + \delta e_2 \cup e_1\right)$.

To be more precise about the signs in \eqref{eq:prod Sf2} and \eqref{eq:prod Uf}, we give the definition of $S_\beta$ for $\beta \in C^2(T,\ZZ_2)$:
\begin{equation}
S_\beta S_{\beta^\prime} = (-1)^{\int \beta^\prime \cup_1 \beta} S_{\beta+\beta^\prime}
\end{equation}
or equivalently
\begin{equation}
S_\beta = (-1)^{\sum_{f < f^\prime \in \beta} f \cup_1 f^\prime } \prod_{f \in \beta} S_f
\end{equation}
where the order of $f$ in $\beta$ doesn't affect the product due to its property \eqref{eq:Sf commutation relation}. Note that the convention for the product $\prod$ is $\prod_{f \in \{f_1, f_2,..., f_n\} } S_f = S_{f_n} \cdots S_{f_2} S_{f_1}$. We can also define $U_\beta$ in the same way. It can be checked that
\begin{equation}
\begin{split}
U_\beta &= (-1)^{\sum_{f < f^\prime \in \beta} f \cup_1 f^\prime } \prod_{f \in \beta} U_f \\
&= :\prod_{f \in \beta} U_f :
\end{split}
\end{equation}
where $:\cdots:$ is the normal ordering which places all $X$ operators to the left of $Z$ operators. For example, we have
\begin{equation}
U_{\delta e} = (\prod_{f \supset e} X_f) \prod_{f^\prime} Z_{f^\prime}^{\int f^\prime \cup_1 \delta e}. 
\label{eq: U delta e}
\end{equation}
On the other hand, we can show that
\begin{equation}
S_{\delta e} = (-1)^{\int_{w_2} e} Z_{f^\prime}^{\int \delta e \cup_1 f^\prime + f^\prime \cup_1 \delta e}
\label{eq: S delta e}
\end{equation}
where the 1-chain $w_2$ consists of all edges of the triangulation, together with the (02) edge for all ``$+$'' tetrahedra and the (13) edge for all ``$-$'' tetrahedra:
\begin{equation}
w_2=\sum_e e + \sum_{t\in +\text{tetrahedra}} t_{02} + \sum_{t\in -\text{tetrahedra}} t_{13}
\end{equation}
This is exactly the 1-chain representing the second Stiefel-Whitney class \cite{GK2016}. It is a 1-cycle, and therefore exact in a topologically-trivial situation.\footnote{Actually, it is also exact if the space is any oriented 3-manifold, but in this paper we limit ourselves to topologically-trivial lattices.} Thus we can define
\begin{equation}
S^E_\beta = (-1)^{\int_E \beta} S_\beta
\end{equation}
where $E$ is a spin structure (i.e. a 2-chain such that $\partial E = w_2$). The identification of $S^E_{\delta e}$ and $U_{\delta e}$ gives us the gauge constraint \eqref{eq:gauge constraints cup1}. Notice that our bosonization map depends on the choice of a spin structure $E$.

The modified Gauss law looks complicated, but it can be written down more concisely if we describe the spin configurations by a 2-cochain $B\in C^2(T,\ZZ_2)$. Our convention is that $B(f)=1$ if $Z_f=-1$ and $B(f)=0$ if $Z_f=1$. Thus the unconstrained Hilbert space is spanned by vectors $\vert B \rangle $ for all $B$. A 2-form gauge transformation has a 1-cochain $\Lambda$ as a parameter and acts by $B\mapsto B+\delta\Lambda$. 
For a general $\Lambda$, the Gauss law constraint is given by
\begin{equation}
\left(\prod_{f \in \delta \Lambda} X_f \right) \left( \prod_{f^\prime} Z_{f^\prime}^{\int  \delta \Lambda \cup_1 f^\prime } \right) (-1)^{\int \Lambda \cup \delta \Lambda}= 1
\label{eq:general Gauss law}
\end{equation}
This formula is proved by $e \cup \delta e =0$ and induction:
\begin{widetext}
\begin{equation}
\begin{split}
&(\prod_{f_1 \in \delta \Lambda_1} X_{f_1}) (\prod_{f_1^\prime} Z_{f_1^\prime}^{\delta \Lambda_1 \cup_1 f_1^\prime}) (-1)^{\int \Lambda_1 \cup \delta \Lambda_1}
(\prod_{f_2 \in \delta \Lambda_2} X_{f_2}) (\prod_{f_2^\prime} Z_{f_2^\prime}^{\delta \Lambda_2 \cup_1 f_2^\prime}) (-1)^{\int \Lambda_2 \cup \delta \Lambda_2} \\
&=\left(\prod_{f \in \delta (\Lambda_1+\Lambda_2)} X_{f}\right) \left(\prod_{f^\prime} Z_{f^\prime}^{\delta (\Lambda_1+\Lambda_2) \cup_1 f^\prime} \right) (-1)^{\int \Lambda_1 \cup \delta \Lambda_1+\Lambda_2 \cup \delta \Lambda_2} (-1)^{\int \delta \Lambda_1 \cup_1 \delta \Lambda_2} \\
&=\left(\prod_{f \in \delta (\Lambda_1+\Lambda_2)} X_{f} \right) \left(\prod_{f^\prime} Z_{f^\prime}^{\delta (\Lambda_1+\Lambda_2) \cup_1 f^\prime} \right) (-1)^{\int (\Lambda_1+\Lambda_2) \cup \delta (\Lambda_1+\Lambda_2)}
\end{split}
\label{eq:use cup1 2}
\end{equation}
\end{widetext}
where we use the identity $\int \delta \Lambda_1 \cup_1 \delta \Lambda_2 = \int \Lambda_1 \cup \delta \Lambda_2 + \delta \Lambda_2 \cup \Lambda_1$ in the last equality.
Eq. \eqref{eq:general Gauss law} can be rewritten as
\begin{equation}
\left(\prod_{f \in \delta \Lambda} X_f \right) (-1)^{\int \Lambda \cup \delta \Lambda + \delta \Lambda \cup_1 B}= 1.
\label{eq:general Gauss law 2}
\end{equation}
Consider now the following 2-form gauge theory defined on a general triangulated 4D manifold $Y$:
\begin{equation}\label{S4dsimple}
S(B)= \sum_t |\delta B (t)| + i \pi \int_Y (B \cup B + B \cup_1 \delta B).
\end{equation}
Here $B\in C^2(Y,\ZZ_2)$, and the gauge symmetry acts by $B \rightarrow B + \delta \Lambda$. The second term is the generalized Steenrod square term defined in \cite{LZW18}. The action is gauge-invariant up to a boundary term:
\begin{equation}
S(B+\delta \Lambda) - S(B) = \int_{\partial Y} ( \Lambda \cup \delta \Lambda + \delta \Lambda \cup_1 B).
\label{eq:Ssimple boundary}
\end{equation}
This boundary term determines the Gauss law for the wave-function $\Psi(B)$ on the spatial slice $X=\partial Y$:
\begin{equation}
\Psi(B+\delta \Lambda) = (-1)^{\omega(\Lambda,B)} \Psi (B)
\end{equation}
where $\omega(\Lambda,B) = \int_X ( \Lambda \cup \delta \Lambda + \delta \Lambda \cup_1 B)$. The Gauss law is the same as the gauge constraint \eqref{eq:general Gauss law 2}. In the following section we use this observation to construct a 4D lattice action for  particular Hamiltonian gauge theories with the modified Gauss law.

\section{Euclidean 3+1D gauge theories with fermionic duals}

In this section we write down Euclidean formulations of some of the gauge theories which are dual to free fermions. We will make use of cup products, and thus will assume that the 3d space is triangulated. Accordingly, the 3+1d lattice will be the product of the 3d triangulation and discrete time. As explained in the Appendix, (higher) cup products can also be defined on the 3d cubic lattice, thus similar considerations can be used to find the Euclidean formulation of gauge theories constructed in Section \ref{sec:square}.

Consider the simplest gauge-invariant Hamiltonian compatible with the modified Gauss law:
\begin{equation}
H= -A \sum_f U_f - B \sum_t W_t.
\end{equation}
The gauge constraint is
\begin{equation}
G_e \equiv \left(\prod_{f \supset e} X_f \right) \prod_{f^\prime} Z_{f^\prime}^{\int  \delta e \cup_1 f^\prime } = 1.
\end{equation}
The partition function is
\begin{equation}
\begin{split}
Z&=\text{Tr } e^{-\beta H} =  \text{Tr } T^M \\
\end{split}
\label{eq:partition function}
\end{equation}
where $T$ is the transfer matrix defined as
\begin{equation}
T= \left(\prod_e \delta_{G_e, 1 }\right) e^{-\delta \tau H} .
\end{equation}
The first factor arises from the gauge constraints on the Hilbert space. For calculation purposes, we can rewrite it as
\begin{equation}
    \begin{split}
        &\delta_{G_e, 1} = \frac{1}{2}(1+G_e) \\
        &= \frac{1}{2} \sum_{\lambda_e = \pm 1}  (-1)^{ \frac{1-\lambda_e}{2} \sum_{f^\prime \in \text{NE}(e)} \frac{1-Z_{f^\prime}}{2}} \cdot   (-1)^{\frac{1-\lambda_e}{2} \sum_{f \supset e} \frac{1-X_f}{2}}  
    \end{split}
\label{eq:trick 1}
\end{equation}
with $\text{NE}(e)\equiv \{f | \int \delta e \cup_1 f =1\}$.
Define $| m(\tau) \rangle = | \{S_f\} \rangle$ as the configuration of spins (in $Z_f$ basis). Then the matrix elements of $T$ are
\begin{equation}
\langle m^\prime (\tau + \delta \tau) | T | m(\tau) \rangle = \langle m^\prime(\tau + \delta \tau) | \left(\prod_v \delta_{G_e, 1} \right) e^{-\delta \tau H}  | m(\tau) \rangle 
\label{eq:T matrix element}
\end{equation}
Next we need to use an identity
\begin{equation}
\langle {S^z}^\prime | f(\sigma^x,\sigma^z) |S^z \rangle = \frac{1}{2} \sum_{S^x = \pm 1} f(S^x,S^z) (-1)^{\frac{1-S^x}{2} (\frac{1-{S^z}^\prime}{2}+\frac{1-S^z}{2})}
\end{equation}
where we assume that $\sigma^x$ is to the right of $\sigma^z$ in the function $f(\sigma^x,\sigma^z)$.
Plugging this into \eqref{eq:T matrix element}, we get
\begin{widetext}
\begin{equation}
\begin{split}
 & \langle m^\prime(\tau + \delta \tau) | \left(\prod_e \delta_{G_e, 1} \right) e^{-\delta \tau H}  | m(\tau) \rangle  \\
&= \langle m^\prime(\tau + \delta \tau) | \left(\prod_e \delta_{G_e, 1}\right) \left(\prod_f \sum_{S^x_f =\pm 1} | S^x_f \rangle \langle S^x_f | \right) e^{-\delta \tau H}  | m(\tau) \rangle \\
 &\propto \left[  \sum_{\lambda_e = \pm 1}  (-1)^{\frac{1-\lambda_e}{2} \left(\sum_{f_2 \supset e} \frac{1-S^x_{f_2}}{2} + \sum_{f_3 \in \text{NE}(e)} \frac{1-{S^z_{f_3}}^\prime}{2} \right)}
 (-1)^{\sum_{\lambda_e,\lambda_{e^\prime}=-1} \int e \cup \delta e^\prime} \right]  \\
& \left[ \prod_f \sum_{S^x_f = \pm1} (-1)^{\frac{1-S^x_f}{2} \left(\frac{1-{S^z_f}^\prime}{2} + \frac{1-S^z_f}{2} \right)} e^{ A \delta \tau S^x_e \prod_{f_1 \in \Delta (f)} S^z_{f_1} } \right] \left(\prod_t e^{ B \delta \tau \prod_{f_4 \subset t} S^z_{f_4}} \right) 
\end{split}
\end{equation}
\end{widetext}
where $\Delta(f) \equiv \{ f^\prime | f^\prime \cup_1 f =1 \}$ and the term $(-1)^{\sum_{\lambda_e,\lambda_{e^\prime}=-1} \int e \cup \delta e^\prime}$ comes from pushing all $X_f$ operators to the right, which is the same as the last factor of Eq. \eqref{eq:general Gauss law}. This term can be expressed as
\begin{equation}
i \pi \sum_i \int a_i \cup \delta a_i
\end{equation}
if we define $a_i$ as 1-cochain on the $i^{\rm th}$ layer with $a_i(e)=1$ for $\lambda_e = -1$ and $a_i(e)=0$ for $\lambda_e = 1$. We can interpret $a_i$ at the $i^{\rm th}$ layer as a 2-cochain which lives on the ``temporal'' faces between the $i^{\rm th}$ and $i+1^{\rm th}$ layers.

After extracting this factor, the remaining terms are
\begin{widetext}
\begin{equation}
\begin{split}
&\sum_{\lambda_e = \pm 1}  \left( \prod_e  (-1)^{\frac{1-\lambda_e}{2} \sum_{f_3 \subset \text{NE}(e)} \frac{1-{S^z_{f_3}}^\prime}{2} } \right) \left(\prod_t e^{ B \delta \tau \prod_{f_4 \subset t} S^z_{f_4}}\right) \\
& \,\,\,\,\,\,\,\,  \left[ \prod_f \sum_{S^x_f = \pm1} (-1)^{\frac{1-S^x_f}{2} \left(\frac{1-{S^z_f}^\prime}{2} + \frac{1-S^z_f}{2}+\sum_{e \subset f} \frac{1-\lambda_e}{2} \right) }  \cdot e^{ A \delta \tau S^x_e \prod_{f_1 \in \Delta (f)} S^z_{f_1} }  \right] \\
&= \sum_{\lambda_e = \pm 1} \left( \prod_e    (-1)^{\frac{1-\lambda_e}{2} \sum_{f_3 \subset \text{NE}(e)} \frac{1-{S^z_{f_3}}^\prime}{2} } \right) \left(\prod_t e^{ B \delta \tau \prod_{f_4 \subset t} S^z_{f_4}}\right) \\
& \,\,\,\,\,\,\,\,  \left[ \prod_e (e^{ A \delta \tau \prod_{f_1 \in \Delta (f)} S^z_{f_1} } + e^{ -A \delta \tau \prod_{f_1 \in \Delta (f)} S^z_{f_1} } {S^z_f}^\prime S^z_f \prod_{e \subset f} \lambda_{e}  ) \right] \\
&\sim \sum_{\lambda_e = \pm 1} \left( \prod_e    (-1)^{\frac{1-\lambda_e}{2} \sum_{f_3 \subset \text{NE}(e)} \frac{1-{S^z_{f_3}}^\prime}{2} } \right) \left(\prod_t e^{ B \delta \tau \prod_{f_4 \subset t} S^z_{f_4}}\right) \\
& \,\,\,\,\,\,\,\,  \left[ \prod_f e^{J {S^z_f}^\prime S^z_f \prod_{e \subset f} \lambda_{e} } (-1)^{\left(\sum_{f_1 \in \Delta(f)} \frac{1-S^z_{f_1}}{2} \right) \left(\frac{1-{S^z_f}^\prime}{2} + \frac{1-S^z_f }{2} + \sum_{e \subset f} \frac{1-\lambda_{e}}{2} \right)} \right] \\
&= \sum_{\lambda_e = \pm 1} (-1)^{ \sum_e \left( \frac{1-\lambda_e}{2} \sum_{f_3 \subset \text{NE}(e)} \frac{1-{S^z_{f_3}}^\prime}{2} \right) + \sum_f \left(\sum_{f_1 \in \Delta(f)} \frac{1-S^z_{f_1}}{2} \right) \left(\frac{1-{S^z_f}^\prime}{2} + \frac{1-S^z_f }{2} + \sum_{e_1 \subset f} \frac{1-\lambda_{e_1}}{2} \right)}  \\
& \,\,\,\,\,\,\,\, e^{J \sum_f {S^z_f}^\prime S^z_f \prod_{e_1 \subset f} \lambda_{e_1} + B \delta \tau \sum_t \prod_{f_4 \subset t} S^z_{f_4}}
\end{split}
\label{eq:Path integral}
\end{equation}
\end{widetext}
where $\tanh J = e^{-2 A \delta \tau}$. The last line is the usual action for a 4D $\ZZ_2$ gauge theory except for some sign factors. We regard these factors  as coming from a topological action $S_{\text{top}}$.
From \eqref{eq:Path integral}, we see that $S_{\text{top}}$ contains
\begin{widetext}
\begin{equation}
\begin{split}
i \pi \biggl[ &\sum_e \frac{1-\lambda_e}{2} \sum_{f_3 \subset \text{NE}(e)} \frac{1-{S^z_{f_3}}^\prime}{2} + \sum_f \left( \sum_{f_1 \in \Delta(f)} \frac{1-S^z_{f_1}}{2} \right) \left(\frac{1-{S^z_f}^\prime}{2} + \frac{1-S^z_f }{2} + \sum_{e_1 \subset f} \frac{1-\lambda_{e_1}}{2} \right) \biggl].
\end{split}
\end{equation}
\end{widetext}
The first term is
\begin{equation}
    \begin{split}
        &\sum_e \frac{1-\lambda_e}{2} \sum_{f\supset e} \sum_{f \cup_1 f_3 =1} \frac{1-{S^z_{f_3}}^\prime}{2}\\
        &=\sum_f \Big(\sum_{e \subset f} \frac{1-\lambda_e}{2}\Big) \Big(\sum_{f \cup_1 f_3 =1}\frac{1-{S^z_{f_3}}^\prime}{2}\Big)
    \end{split}
\end{equation}
which is equal to $\int \delta a_i \cup_1 b_{i+1}$ if we define $b_i$ as a 2-cochain on the $i$th layer with $b_i(f)=\frac{1-S_f}{2}$. The second term is
\begin{equation}
\begin{split}
\sum_f &\Big(\sum_{f_1 | \int f_1 \cup_1 f =1} \frac{1-S^z_{f_1}}{2}\Big) \\
& ~~\cdot \Big(\frac{1-{S^z_f}^\prime}{2} + \frac{1-S^z_f }{2} + \sum_{e_1 \subset f} \frac{1-\lambda_{e_1}}{2} \Big)
\end{split}
\end{equation}
which is $\int b_i \cup_1 (b_i + b_{i+1} + \delta a_i)$. Collecting all terms, we get
\begin{equation}
\begin{split}
S_{\text{top}} (\{a_i\},\{b_i\}) =& i \pi \sum_i \int a_i \cup \delta a_i + \delta a_i \cup_1 b_{i+1} \\
&+ b_i \cup_1 (b_i + b_{i+1} + \delta a_i).
\end{split}
\label{eq:Stop}
\end{equation}
The usual term $e^{J \sum_f {S^z_f}^\prime S^z_f \prod_{e_1 \subset f} \lambda_{e_1} + B \delta \tau \sum_T \prod_{f_4 \subset T} S^z_{f_4}}$ can be written as the exponential of (up to an unimportant constant)
\begin{equation}
\begin{split}
&S_{\text{4D gauge}}(\{a_i\},\{b_i\})\\
&= \sum_i \Big( -2J \sum_f |b_i(f)+b_{i+1}(f)+\delta a_i (f)| \\
& ~~~~ ~~~~ ~~ -2B \delta \tau \sum_t |\delta b_i (t)|\Big)
\end{split}
\label{eq:S4D}
\end{equation}
where $|\cdots|$ gives the argument's parity $0$ or $1$.
Combining \eqref{eq:Stop} and \eqref{eq:S4D}, the Euclidean action becomes (up to an additive constant)
\begin{equation}
S(\{a_i\},\{b_i\})= S_{\text{top}} (\{a_i\},\{b_i\}) + S_{\text{4D gauge}} (\{a_i\},\{b_i\}) ,
\label{eq:S4dtotal}
\end{equation}
which is analogous to action \eqref{S4dsimple} defined by generalized Steenrod square..
This action is gauge-invariant under gauge transformations
\begin{equation}
b_i \rightarrow b_i + \delta \lambda_i, \,\, a_i \rightarrow a_i + \delta \mu_i + \lambda_i + \lambda_{i+1},
\label{eq:gauge transformation}
\end{equation}
where $\lambda_i$ are arbitrary 1-cochains and $\mu_i$ are arbitrary 0-cochains. Indeed, the change in the action is
\begin{widetext}
\begin{equation}
\begin{split}
\frac{\Delta S_{\text{top}}}{(i \pi)} =& \sum_i \int (a_i + \cancelto{0}{\delta \mu_i} + \lambda_i +  \lambda_{i+1}) \cup (\delta \lambda_i + \delta \lambda_{i+1}) \\
& +(\cancelto{0}{\delta \mu_i} + \lambda_i + \lambda_{i+1}) \cup \delta a_i +(\textcolor{red}{\delta \lambda_i+ \delta \lambda_{i+1}} ) \cup_1 b_{i+1} + \delta a_i \cup_1 \delta \lambda_{i+1}  \\
& +(\delta \lambda_i + \cancelto{0}{\delta \lambda_{i+1}}) \cup_1 \delta \lambda_{i+1} + \delta \lambda_i \cup_1 (\textcolor{red}{b_i+ b_{i+1}}  + \delta a_i) \\
=& \sum_i \int \textcolor{blue}{a_i \cup (\delta \lambda_i + \delta \lambda_{i+1})} + ( \lambda_i + \lambda_{i+1}) \cup (\delta \lambda_i + \delta \lambda_{i+1}) \\
&+\textcolor{cyan}{(\lambda_i + \lambda_{i+1}) \cup \delta a_i} + \textcolor{blue}{a_i \cup \delta \lambda_{i+1}} + \textcolor{cyan}{\delta \lambda_{i+1} \cup a_i} + \lambda_i \cup \delta \lambda_{i+1} + \delta \lambda_{i+1} \cup \lambda_i \\
&+ \textcolor{blue}{a_i \cup \delta \lambda_i} + \textcolor{cyan}{\delta \lambda_i \cup a_i} \\
=& \sum_i \int \lambda_i \cup \delta  \lambda_i + \lambda_{i+1} \cup \delta \lambda_{i+1} = 0
\end{split}
\label{eq:use cup1 3}
\end{equation}
\end{widetext}
where the terms with the same colors cancel out. In the above computation we assumed periodic time, so that there are no boundary terms. If we do not identify time periodically, the variation is a boundary term
\begin{equation}
\int (\lambda_0 \cup \lambda_0 + \delta \lambda_0 \cup_1 b_0) + (\lambda_N \cup \lambda_N + \delta \lambda_N \cup_1 b_N),
\end{equation}
which is the same as the boundary term \eqref{eq:Ssimple boundary} in the previous section.

We can also check that the action is invariant under a 2-form global symmetry
\begin{equation}
B \rightarrow B + \beta
\end{equation}
where a closed 2-cochain $\beta$ can be represented by 2-cochains $\beta_i$ (one for each time slice) and 1-cochains $\alpha_i$ satisfying $\beta_i + \beta_{i+1} + \delta \alpha_i = 0$. Using a gauge transformation \eqref{eq:gauge transformation} with
\begin{equation}
\lambda_i = \sum^{i-1}_{j=0} \alpha_j, \,\,\,\, \mu_i=0
\end{equation}
for $i=0,1,...,N-1$, we can see that $\beta^\prime_i=\beta_0$, which is independent of $i$, and $\alpha^\prime_{N-1}=\sum^{N-1}_{j=0} \alpha_j$ with other $\alpha^\prime_i=0$. Notice that $\alpha^\prime_{N-1}$ is closed since $\beta^\prime_i=\beta^\prime_{i+1}$. Under this 2-form symmetry transformation $\beta^\prime$, the action changes by
\begin{equation}
\begin{split}
\frac{\Delta S_{\text{top}}}{(i \pi)} =& \int \cancelto{0}{\alpha^\prime_{N-1} \cup \delta a_{N-1}} + \sum_i \delta a_i \cup_1 \beta_0 \\
&+ \beta_0 \cup_1 (\cancelto{0}{\sum_i b_i + b_{i+1}}) + \sum_i \beta_0 \cup_1 \delta a_i \\
=& \sum_i \int a_i \cup \beta_0 + \beta_0 \cup a_i + \beta_0 \cup a_i + a_i \cup \beta_0 = 0.
\end{split}
\end{equation}
Thus the action is invariant under a global 2-form symmetry, as expected.

\section{Gauging fermion parity}

We have shown that a lattice fermionic system in 3d is dual to a bosonic spin system with the Gauss law constraints. In this section we show how to get rid of the constraints at the expense of coupling fermions to a $\ZZ_2$ gauge field.

Our bosonization map is
\begin{equation}
\begin{split}
(-1)^{F_t} = - i \g_t \g^\prime_t &\longleftrightarrow W_t \equiv \prod_{f \subset t} Z_f \\
S^E_f = (-1)^{\int_E f} (i \g_{L(f)} \g^\prime_{R(f)})  &\longleftrightarrow U_f \equiv X_f \prod_{f^\prime} Z_{f^\prime}^{\int f^\prime \cup_1 f}
\end{split}
\label{eq: original bosonization map}
\end{equation}
with gauge constraints
\begin{equation}
\left(\prod_{f \supset e} X_f \right) \prod_{f^\prime} Z_{f^\prime}^{\int  \delta e \cup_1 f^\prime } = 1.
\end{equation}
Now, we introduce new $\ZZ_2$ fields (spins), with operators $\tilde{X}$, $\tilde{Y}$, and $\tilde{Z}$, which live on faces and couple to fermions via a Gauss law constraint
\begin{equation}\label{eq:Gauss new}
(-1)^{F_t} = \prod_{f \subset t} \tilde{Z}_f.
\end{equation}
The fermionic hopping operator must be modified to
\begin{equation}S^E_f = (-1)^{\int_E f} (i \g_{L(f)} \g^\prime_{R(f)}) \tilde{X}_f
\end{equation}
in order to commute with the Gauss law constraint (\ref{eq:Gauss new}). 
The bosonization map becomes
\begin{equation}
\begin{split}
- i \g_t \g^\prime_t = \prod_{f \subset t} \tilde{Z}_f &\longleftrightarrow W_t \equiv \prod_{f \subset t} Z_f \\
S^E_f = (-1)^{\int_E f} (i \g_{L(f)} \g^\prime_{R(f)}) \tilde{X}_f &\longleftrightarrow U_f \equiv X_f \prod_{f^\prime} Z_{f^\prime}^{\int f^\prime \cup_1 f}
\end{split}
\label{eq: new bosonization map 1}
\end{equation}
and, similar to \eqref{eq: U delta e} and \eqref{eq: S delta e}, the identification of $U_{\delta e}$ and $S_{\delta e}$ gives
\begin{equation}
\prod_{f \supset e} \tilde{X}_f \longleftrightarrow \left(\prod_{f \supset e} X_f \right) \prod_{f^\prime} Z_{f^\prime}^{\int  \delta e \cup_1 f^\prime }
\label{eq: new bosonization map 2}
\end{equation}
The equations \eqref{eq: new bosonization map 1} and \eqref{eq: new bosonization map 2} define a bosonization map for fermions coupled to a dynamical $\ZZ_2$ gauge field. In this case, their is no constraint on the bosonic variables.

We can apply this modified boson/fermion map to a $\ZZ_2$ version of the Levin-Wen rotor model \cite{LW2006} on general triangulation:

\begin{equation}
H = - \sum_t Q_t - \sum_e B_e
\label{eq:generalized Levin Wen}
\end{equation}
with
\begin{equation}
\begin{split}
Q_t &= \prod_{f \subset t} Z_f \\
B_e &= \prod_{f \supset e} \left(X_f \prod_{f^\prime} Z_{f^\prime}^{\int f \cup_1 f^\prime} \right) \\
&= \left(\prod_{f \supset e} X_f \right) \prod_{f^\prime} Z_{f^\prime}^{\int \delta e \cup_1 f^\prime}
\end{split}
\end{equation}
Since $Q_t$ and $B_e$ are just $W_t$ and $(\prod_{f \supset e} X_f) \prod_{f^\prime} Z_{f^\prime}^{\int  \delta e \cup_1 f^\prime }$, the above bosonic model  is equivalent to a model of a $\ZZ_2$ gauge field coupled to fermions and a Hamiltonian
\begin{equation}
H = - \sum_t \prod_{f \subset t} \tilde{Z}_f - \sum_e \prod_{f \supset e} \tilde{X}_f.
\end{equation}
The fermions are static, since the above Hamiltonian does not include fermionic hopping terms. The only interaction between the fermions and the gauge field is via the Gauss law constraint
\begin{equation}
\prod_{f \subset t} \tilde{Z}_f=(-1)^{F_t}.
\end{equation}
Thus a complicated model bosonic model is mapped to a simple $\ZZ_2$ lattice gauge theory coupled to static fermions. 

As another application of the modified bosonization map, consider again the bosonic gauge theory on a cubic lattice with the Hamiltonian \eqref{eq:bosonic Creutz model} 
\begin{equation}
H = - \frac{t}{2} \sum_{i=x,y,z} \sum_{f_i} s_i(L(f_i)) U_{f_i} (1- W_{L(f_i)} W_{R(f_i)})
\label{eq:bosonic Creutz model without constraint}
\end{equation}
and a gauge constraint \eqref{eq:cubic gauge constraint}. This constrained model is dual a model of free fermions with Dirac cones. After coupling the fermions to a $\ZZ_2$ gauge field and applying the modified map, we find that the bosonic model \eqref{eq:bosonic Creutz model without constraint} without any gauge constraints is equivalent to a fermionic model with the Hamiltonian
\begin{equation}
\begin{split}
H = -t \sum_{\vec{r}} \Big(& s_x(\vec{r})\tilde{X}_x(\vec{r}) c^\dagger_{\vec{r}+\hat{x}} c_{\vec{r}}  + s_y(\vec{r})\tilde{X}_y(\vec{r}) c^\dagger_{\vec{r}+\hat{y}} c_{\vec{r}} \\
&+ s_z(\vec{r}) \tilde{X}_z(\vec{r})c^\dagger_{\vec{r}+\hat{z}} c_{\vec{r}} + \text{h.c.} \Big)
\end{split}
\label{eq:modified Creutz model}
\end{equation}
with $(-1)^{c^\dagger_t c_t} = \prod_{f \subset t} \tilde{Z}_f$. The operators $ \tilde{W}_e \equiv \prod_{f \supset e} \tilde{X}_f$ commute with the Hamiltonian, so we can project the Hilbert space into sectors with fixed $\tilde{W}_e$ ($\tilde{W}_e$ is arbitrary $\pm 1$ as long as it satisfies $\prod_{e \supset v} \tilde{W}_e =1$). In the sector $\tilde{W}_e=1$ for all $e$, the Hamiltonian \eqref{eq:modified Creutz model} returns to \eqref{eq: Creutz Hamiltonian}. The model of unconstrained spins with the Hamiltonian \eqref{eq:bosonic Creutz model without constraint} thus can be regarded as a 3d analog of Kitaev's honeycomb model.

\section{Conclusions}

In this paper we constructed a bosonization map for an arbitrary fermionic system on a 3d lattice. The lattice can be either a cubic one or a triangulation. The dual bosonic system is a 2-form gauge theory and thus has local constraints (the modified Gauss law). While we did not emphasize this point in the paper, the form of the constraints is largely determined by requiring the system to have a 2-form $\ZZ_2$ symmetry with a particular 't Hooft anomaly. As explained in the end of section 2, another way to understand the constraints is to note that they arise from a 4D 2-form gauge theory with a ``topological'' term (generalized Steenrod square) in the action
\begin{equation}
S_{top}=i\pi \int_Y B\cup B + B \cup_1 \delta B,
\end{equation}
where $B$ is a 2-form $\ZZ_2$ gauge field (i.e. a 2-cochain with values in $\ZZ_2$). This action is invariant under a gauge symmetry $B\mapsto B+\delta\lambda$, where $\lambda$ is a 1-cochain, up to a non-trivial boundary term, and it is this boundary term which leads to a modified Gauss law. 

One can get rid of the constraint on the bosonic side at the expense of coupling the fermions to a $\ZZ_2$ gauge field (i.e. by gauging the fermion parity). We used this observation to construct a model with spins and no constraints which is dual to a model of free fermions coupled to a static gauge field, and thus is soluble.

The simplest Euclidean 4D 2-form gauge theory which leads to the correct form of the Gauss law for the wave-functions has the action (\ref{S4dsimple}). It is very likely that this model is dual to a model of free fermions for any triangulated 4D manifold $Y$. It would be very interesting to prove this. Our methods are insufficient here, since they are tied to the Hamiltonian formalism, while (\ref{S4dsimple}) makes sense only on a 4D triangulation, but not on a 3d triangulation times discrete time, and thus is intrinsically Euclidean. In this paper, instead of attacking this problem head-on, we showed that a complicated-looking 2-form gauge theory with an action (\ref{eq:S4dtotal}) is dual to a theory of free fermions. This Euclidean theory leads to the same Gauss law as (\ref{S4dsimple}), but has the advantage that it is defined on a 3d triangulation times discrete time, and thus can be analyzed by our methods.

\appendix
\section{(Higher) cup products on a triangulation and a cubic lattice} \label{sec:cup product}

In the case of a general triangulation, our bosonization procedure is based on the properties of the cup product $\cup$ and the higher cup product $\cup_1$. These mathematical operations have been defined by Steenrod \cite{Steenrod} (see also Appendix B in~\cite{KapSei} for a review) for an arbitrary simplicial complex, but not for a cubic lattice. In this section, we will describe a version of these definitions for the cubic lattice and check that the usual properties of these products hold. 

For a simplicial complex, the cup product of cochains $\cup$  is defined as \cite{Hatcher}
\begin{equation}
\begin{split}
&[A_p\cup B_p](0, 1, \cdots, p+q) \\
&= A_p(0,1,\cdots p) B_q (p,p+1,\cdots,p+q),
\end{split}
\end{equation}
while the higher cup product $\cup_1$  is defined as \cite{Steenrod,KapSei}
\begin{equation}
\begin{split}
&[ A_p \cup_1 B_q] (0,\cdots , p+q-1) \\
&= \sum_{i_0} A(0,\cdots,i_0,i_0+q,\cdots,p+q-1) B(i_0,\cdots,i_0+q).
\end{split}
\end{equation}
Here $A_p$ and $B_q$ are arbitrary $p$-cochain and $q$-cochains with values in $\ZZ_2$. We will limit ourselves to the case of $\ZZ_2$ valued cochains, since this is all we need in this paper.

\begin{figure}[htb]
\centering
\resizebox{8cm}{!}{%
\begin{tikzpicture}
\draw[thick] (-0.5,0) -- (2.5,0);\draw[thick] (3,1) -- (3.5,1);\draw[thick] (0.5,2.5) -- (3.5,2.5);\draw[thick] (-0.5,1.5) -- (2.5,1.5);

\draw[thick] (-0.3,-0.3) -- (0,0);\draw[thick] (1.7,-0.3) -- (3.3,1.3);\draw[thick] (-0.3,1.2) -- (1.3,2.8);\draw[thick] (1.7,1.2) -- (3.3,2.8);

\draw[thick] (0,-0.5) -- (0,2);\draw[thick] (1,2.5) -- (1,3);\draw[thick] (3,0.5) -- (3,3);\draw[thick] (2,-0.5) -- (2,2);

\draw[thick,dashed] (0,0) -- (1.3,1.3);\draw[thick,dashed] (0.5,1) -- (3,1);\draw[thick,dashed] (1,0.5) -- (1,2.5);

\draw[->][thick] (0,0) -- (1,0);\draw[->][thick] (0,1.5) -- (1.2,1.5);\draw[->][thick,dashed] (1,1) -- (1.9,1);\draw[->][thick] (1,2.5) -- (2,2.5);
\draw[->][thick] (0,0) -- (0,0.75);\draw[->][thick] (2,0) -- (2,0.75);\draw[->][thick,dashed] (1,1) -- (1,1.8);\draw[->][thick] (3,1) -- (3,1.75);
\draw[->][thick,dashed] (0,0) -- (0.5,0.5);\draw[->][thick] (2,0) -- (2.5,0.5);\draw[->][thick] (0,1.5) -- (0.5,2);\draw[->][thick] (2,1.5) -- (2.5,2);

\draw[->][thick] (-2,1) -- (-1,1);\draw[->][thick] (-2,1) -- (-2,2);\draw[->][thick] (-2,1) -- (-1.4,1.6);
\draw (-0.8,1) node{x};\draw (-1.2,1.8) node{y};\draw (-2,2.2) node{z};

\filldraw[black] (0,0) circle(1pt) node[anchor=north west] {0};
\filldraw[black] (2,0) circle(1pt) node[anchor=north west] {1};
\filldraw[black] (0,1.5) circle(1pt) node[anchor=south east] {2};
\filldraw[black] (1,1) circle(1pt) node[anchor=north west] {3};
\filldraw[black] (3,1) circle(1pt) node[anchor=north west] {4};
\filldraw[black] (2,1.5) circle(1pt) node[anchor=south east] {5};
\filldraw[black] (1,2.5) circle(1pt) node[anchor=south east] {6};
\filldraw[black] (3,2.5) circle(1pt) node[anchor=south east] {7};

\draw[thick] (4.5,1)--(6.5,1);\draw[thick] (5.5,0)--(5.5,2);\draw[thick] (4.9,0.4)--(6.1,1.6);
\draw[->][thick] (5.5,1)--(5,1);\draw[->][thick] (6.5,1)--(6,1);
\draw[->][thick] (4.9,0.4)--(5.2,0.7);\draw[->][thick] (5.5,1)--(5.8,1.3);
\draw[->][thick] (5.5,2)--(5.5,1.5);\draw[->][thick] (5.5,1)--(5.5,0.5);
\draw (5.5,2)  node[anchor=south] {U};
\draw (5.5,0)  node[anchor=north] {D};
\draw (6.5,1)  node[anchor=west] {R};
\draw (4.5,1)  node[anchor=east] {L};
\draw (4.9,0.4)  node[anchor=north east] {F};
\draw (6.1,1.6)  node[anchor=south west] {B};

\end{tikzpicture}
}
\caption{ There are six faces for each cube $c$. U,D,F,B,L,R stand for faces on direction "up","down","front","back","left","right". We assign the face U, F, R to be inward and D, B, L to be outward. The $\cup_1$ product on two 2-cochain is defined by $\beta \cup_1 \beta^\prime (c) = \beta (L)\beta^\prime(B) + \beta (L)\beta^\prime(D) + \beta (B)\beta^\prime(D) + \beta (U)\beta^\prime(F) + \beta (U)\beta^\prime(R) + \beta (F)\beta^\prime(R)$}
\label{fig:cubic cup}
\end{figure}
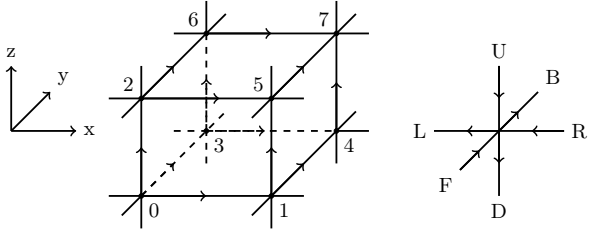

To generalize these formulas to the cubic lattice, we first develop an intuition for the cup product $\cup$. On a triangle $\Delta_{012}$, the usual cup product for two 1-cochains $\lambda$ and $\lambda^\prime$ is
\begin{equation}
\lambda \cup \lambda^\prime (012)= \lambda(01) \lambda^\prime (12).
\end{equation}
We can think of it as starting from vertex $0$, passing through edges $01$ and $12$ consecutively, and ending at vertex $2$, all the while following the orientation of the edges. Following the same logic, it is intuitive to define the cup product on a square $\Box_{0134}$ (the bottom face in Fig. \ref{fig:cubic cup}) as
\begin{equation}
\lambda \cup \lambda^\prime (0134)= \lambda(01) \lambda^\prime (14) + \lambda(03) \lambda^\prime (34).
\end{equation}
The two terms come from two oriented paths from vertex 0 to vertex 4.
If $\lambda$ and $\beta$ are a 1-cochain and a 2-cochain, the usual cup product is
\begin{equation}
\begin{split}
\lambda \cup \beta (0123) &= \lambda (01) \beta (123) \\
\beta \cup \lambda (0123) &= \beta (012) \lambda (23).
\end{split}
\end{equation}
On the cubic lattice, the corresponding cup products are defined as follows:
\begin{equation}
\begin{split}
&\lambda \cup \beta (c) \\
&= \lambda (01) \beta (\Box_{1457}) + \lambda (02) \beta (\Box_{2567}) + \lambda (03) \beta (\Box_{3467}) \\
&\beta \cup \lambda (c) \\
&= \beta (\Box_{0236}) \lambda (67) + \beta (\Box_{0125}) \lambda (57) + \beta (\Box_{0134}) \lambda (47)
\end{split}
\end{equation}
where $c$ is a cube whose vertices are labeled in Fig. \ref{fig:cubic cup}. For a cup product involving 0-cochains, the definition is trivial:
\begin{equation}
\begin{split}
v \cup \beta (\Box_{0134}) &= v(0) \beta (\Box_{0134}) \\
\beta \cup v (\Box_{0134}) &=  \beta (\Box_{0134}) v(4) \\
v \cup \lambda (01) &= v(0) \lambda (01) \\
\lambda \cup v (01) &= \lambda (01) v(1).
\end{split}
\end{equation}
With the above definitions, it can be checked that the following equalities hold for cubic cochains of degrees $0$, $1$ and $2$:
\begin{equation}
\begin{split}
e_1 \cup \delta e_2 &= \delta e_1 \cup e_2 + \delta (e_1 \cup e_2) \\
v \cup \delta f &= \delta v \cup f + \delta (v \cup f).
\end{split}
\end{equation}

The next step is to define the $\cup_1$ product on the cubic lattice. It need not satisfy all the properties that $\cup_1$ has on a triangulation. The only properties of $\cup_1$ that we need are the anti-commutativity for faces with the same direction and the identity we used in \eqref{eq:use cup1 1}, \eqref{eq:use cup1 2}, and \eqref{eq:use cup1 3}:
\begin{equation}
\int e_1 \cup \delta e_2 + \delta e_2 \cup e_1 = \int \delta e_1 \cup_1 \delta e_2 \,\, (\text{mod } 2).
\label{eq: cup1 need to satisfy}
\end{equation}
Therefore, we only need to define $\cup_1$ product for two 2-cochains so that it satisfies  \eqref{eq: cup1 need to satisfy}. Our convention for $\cup_1$ is shown in Fig. \ref{fig:cubic cup}:
\begin{equation}
\begin{split}
\beta \cup_1 \beta^\prime (c) =& \,\beta (L)\beta^\prime(B) + \beta (L)\beta^\prime(D) + \beta (B)\beta^\prime(D) \\
&+ \beta (U)\beta^\prime(F) + \beta (U)\beta^\prime(R) + \beta (F)\beta^\prime(R)
\end{split}
\end{equation}

Once the $\cup$ and $\cup_1$ products are defined on the cubic lattice, the bosonization procedure on a general triangulation can be applied to the cubic lattice. \eqref{eq:prod Sf1} and \eqref{eq:math way to express prod Sf1} are modified as follows:

\begin{equation}
S_{\delta e} = (-1)^{\sum_{f < f^\prime \in \delta e} f \cup_1 f^\prime } \prod_{f \in \delta e} S_f = \prod_{c | e \in \{01,14,02,47,67,26\} } W_c
\label{eq:prod Sf1 2}
\end{equation}

\begin{equation}
\begin{split}
&Z_{f_1} Z_{f_2} \prod_{f^\prime \subset c} Z_{f^\prime}^{(f_1+f_2) \cup_1 f^\prime + f^\prime \cup_1 (f_1+f_2)} \\
&= 
\begin{cases}
    W_c, & \text{if } e \in \{01,14,02,47,67,26\}  \\
    0,   & \text{otherwise}
\end{cases}
\end{split}
\label{eq:math way to express prod Sf1 2}
\end{equation}
for faces $f_1$ and $f_2$ join at the edge $e$. We implicitly choose $w_2=0$ in \eqref{eq: S delta e}. We can use the $\cup_1$ product defined above to reproduce the fermionic hopping terms defined by framing in Fig. \ref{fig:cube1}.
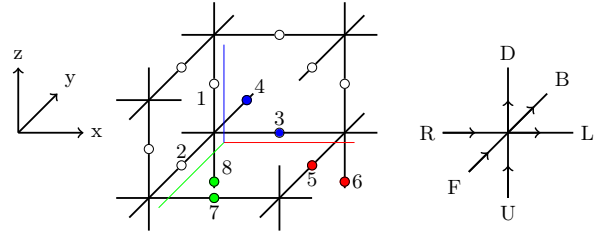
\begin{figure}[htb]
\centering
\resizebox{8cm}{!}{%
\begin{tikzpicture}
\draw[thick] (-0.5,0) -- (2.5,0);\draw[thick] (0.5,1) -- (3.5,1);\draw[thick] (0.5,2.5) -- (3.5,2.5);

\draw[thick] (-0.3,-0.3) -- (1.6,1.6);\draw[thick] (1.7,-0.3) -- (3.3,1.3);\draw[thick] (-0.3,1.2) -- (1.3,2.8);

\draw[thick] (0,-0.5) -- (0,2);\draw[thick] (1,0.15) -- (1,3);\draw[thick] (3,0.15) -- (3,3);

\draw[thick] (-0.5,1.5) -- (0.5,1.5); \draw[thick] (2,-0.5) -- (2,0.5);\draw[thick] (2.3,1.8) -- (3.3,2.8);

\draw[green] (0.15,-0.15) -- (1.15,0.85);
\draw[red] (1.15,0.85) -- (3.15,0.85);
\draw[blue] (1.15,0.85) -- (1.15,2.35);

\draw[->][thick] (-2,1) -- (-1,1);\draw[->][thick] (-2,1) -- (-2,2);\draw[->][thick] (-2,1) -- (-1.4,1.6);
\draw (-0.8,1) node{x};\draw (-1.2,1.8) node{y};\draw (-2,2.2) node{z};

\filldraw[white] (2.5,2) circle (2pt); \draw (2.5,2) circle (2pt);
\filldraw[white] (0.5,0.5) circle (2pt); \draw (0.5,0.5) circle (2pt) node[anchor=south] {2};
\filldraw[white] (0,0.75) circle (2pt); \draw (0,0.75) circle (2pt);
\filldraw[white] (1,1.75) circle (2pt); \draw (1,1.75) circle (2pt) node[anchor=north east] {1};
\filldraw[white] (3,1.75) circle (2pt); \draw (3,1.75) circle (2pt);
\filldraw[white] (0.5,2) circle (2pt); \draw (0.5,2) circle (2pt);
\filldraw[white] (2,2.5) circle (2pt); \draw (2,2.5) circle (2pt);

\filldraw[green] (1,0.25) circle (2pt);\draw (1,0.25) circle (2pt) node[anchor=south west] {8};
\filldraw[green] (1,0) circle (2pt); \draw (1,0) circle (2pt) node[anchor=north] {7};
\filldraw[red] (3,0.25) circle (2pt); \draw (3,0.25) circle (2pt) node[anchor=west] {6};
\filldraw[red] (2.5,0.5) circle (2pt); \draw (2.5,0.5) circle (2pt) node[anchor=north] {5};
\filldraw[blue] (1.5,1.5) circle (2pt); \draw (1.5,1.5) circle (2pt) node[anchor=south west] {4};
\filldraw[blue] (2,1) circle (1.5pt); \draw (2,1) circle (2pt) node[anchor=south] {3};

\draw[thick] (4.5,1)--(6.5,1);\draw[thick] (5.5,0)--(5.5,2);\draw[thick] (4.9,0.4)--(6.1,1.6);
\draw[->][thick] (4.5,1)--(5,1);\draw[->][thick] (5.5,1)--(6,1);
\draw[->][thick] (4.9,0.4)--(5.2,0.7);\draw[->][thick] (5.5,1)--(5.9,1.4);
\draw[->][thick] (5.5,1)--(5.5,1.5);\draw[->][thick] (5.5,0)--(5.5,0.5);
\draw (5.5,2)  node[anchor=south] {D};
\draw (5.5,0)  node[anchor=north] {U};
\draw (6.5,1)  node[anchor=west] {L};
\draw (4.5,1)  node[anchor=east] {R};
\draw (4.9,0.4)  node[anchor=north east] {F};
\draw (6.1,1.6)  node[anchor=south west] {B};

\end{tikzpicture}
}
\caption{(Color online) We rotate the axis U,D,F,B,R,L in Fig. \ref{fig:cubic cup} to match the result in Fig. \ref{fig:cube1}. Notice that the cube above is dual lattice and edges $1,2,3 ...$ in the dual lattice represent faces in the original lattice.}
\label{fig:cup framing}
\end{figure}
The hopping term defined by Eq.~\eqref{eq:hopping defined by cup1} is 
\begin{equation}
U_f = X_f \prod_{f^\prime} Z_{f^\prime}^{\int f^\prime \cup_1 f}.
\label{eq:hopping defined by cup1 2}
\end{equation}
Fig. \ref{fig:cup framing} is dual to Fig. \ref{fig:cubic cup}. Therefore, faces in Fig. \ref{fig:cubic cup} become edges in Fig. \ref{fig:cup framing}. Consider the hopping term along dual edge $3$. On the vertex to the right, it represents the face R. From terms $\beta(F) \beta^\prime(R)$ and $\beta(U) \beta^\prime(R)$, the hopping term contains $Z_5$ (from F) and $Z_6$ (from U). On the vertex to the left, it represents the face L. Since there is no $\beta(D) \beta^\prime(L)$ or $\beta(B) \beta^\prime(L)$ term, it contributes nothing. So we have
\begin{equation}
U_3 = X_3 Z_5 Z_6.
\end{equation}
Similarly, for edge 2, the hopping term has $Z_7$ (from $\beta(L) \beta^\prime(B)$) and $Z_8$ (from $\beta(U) \beta^\prime(F)$)
\begin{equation}
U_2 = X_2 Z_7 Z_8.
\end{equation}
For edge $1$, the hopping term has $Z_3$ (from $\beta(L) \beta^\prime(D)$) and $Z_4$ (from $\beta(B) \beta^\prime(D)$)
\begin{equation}
U_1 = X_1 Z_3 Z_4.
\end{equation}
We get the exact same hopping terms defined by "framing" in Fig. \ref{fig:cube1}.

\end{document}